\documentclass[aps,prl, twocolumn,superscriptaddress,amsfonts,amsmath,amssymb,floatfix,showpacs]{revtex4-1}
\usepackage{graphicx}
\usepackage{subfigure}
\usepackage{hyperref}
\usepackage{ulem}
\usepackage[section]{placeins}
\usepackage{amsmath}
\usepackage{amssymb}
\usepackage{dsfont}
\usepackage{bm}
\usepackage{tikz}
\usepackage{color}
\usepackage{graphics}
\usepackage{float}
\usepackage{longtable}
\usepackage{CJK}
\usepackage{booktabs}
\usepackage{lipsum}

\newcommand{\ket}[1]{|#1\rangle}             
\newcommand{\bra}[1]{\langle#1|}             

\newcommand{\be}{\begin{equation}}
\newcommand{\bea}{\begin{eqnarray}}
\newcommand{\ee}{\end{equation}}
\newcommand{\eea}{\end{eqnarray}}
\newcommand{\ben}{\begin{equation*}}
\newcommand{\bean}{\begin{eqnarray*}}
\newcommand{\een}{\end{equation*}}
\newcommand{\eean}{\end{eqnarray*}}
\newcommand{\ba}{\begin{align}}
\newcommand{\ea}{\end{align}}
\newcommand{\ban}{\begin{align*}}
\newcommand{\ean}{\end{align*}}

\newcommand{\JL}[1]{\textcolor{black}{ #1}}
\newcommand{\mm}[1]{\textcolor{black}{#1}}

\usepackage{mathtools} 

\usepackage{xcolor} 

\usepackage{verbatim}

\begin{document}

\title{Is \mm{high-dimensional photonic} entanglement robust to noise?}
\author{ Feng Zhu}
\affiliation{School of Engineering and Physical Sciences, Heriot-Watt University, Edinburgh, EH14 4AS, UK}

\author{ Max Tyler}
\affiliation{School of Engineering and Physical Sciences, Heriot-Watt University, Edinburgh, EH14 4AS, UK}

\author{Natalia Herrera Valencia}
\affiliation{School of Engineering and Physical Sciences, Heriot-Watt University, Edinburgh, EH14 4AS, UK}

\author{ Mehul Malik}
\affiliation{School of Engineering and Physical Sciences, Heriot-Watt University, Edinburgh, EH14 4AS, UK}
\affiliation{Institute for Quantum Optics and Quantum Information (IQOQI), Austrian Academy of Sciences, Vienna, Austria}

\author{Jonathan Leach}
\affiliation{School of Engineering and Physical Sciences, Heriot-Watt University, Edinburgh, EH14 4AS, UK}

\date{\today}

\begin{abstract}
High-dimensional entangled states are of significant interest in quantum science as they increase the information content per photon and can remain entangled in the presence of significant noise.   We develop the analytical theory and show experimentally that the noise tolerance of high-dimensional entanglement can be significantly increased by modest increases to the size of the Hilbert space. For example, doubling the size of a Hilbert space with local dimension d=300 leads to a reduction of the threshold detector efficiencies required for entanglement certification by two orders of magnitude. This work is developed in the context of spatial entanglement, but it can easily be translated to photonic states entangled in different degrees of freedom. We also demonstrate that knowledge of a single parameter, the signal-to-noise ratio, precisely links measures of entanglement to a range of experimental parameters quantifying noise in a quantum communication system, enabling accurate predictions of its performance.  This work serves to answer a simple question:~``{\it Is high-dimensional photonic entaglement robust to noise?}".  Here we show that the answer is more nuanced than a simple ``yes" or ``no" and involves a complex interplay between the noise characteristics of the state, channel, and detection system.

\end{abstract}
\maketitle

\section{Introduction}

Entanglement is considered one of the most important features of quantum information science and plays a central role in many quantum communication protocols \cite{einstein1935can, horodecki2009quantum, Reid2009}. The strong, non-classical correlations inherent to entanglement allow one to share information between two parties that is secure against the most sophisticated eavesdropping attacks, with information security even provided independent of the devices used \cite{Acin:2007db}. \mm{Entangled photons produced via parametric nonlinear processes are a workhorse for many branches of quantum information science, ranging from fundamental tests of quantum mechanics to entanglement-based quantum communication} \cite{dada2011experimental, giovannetti2011advances, fickler2013real, bornman2019ghost, ringbauer2015measurements, CQKA}. However, photonic entanglement is extremely susceptible to common sources of noise such as channel loss, background counts, multi-photon effects, and imperfect \mm{measurement} devices, which can \mm{make its detection challenging}.

In recent years, high-dimensional entanglement has emerged as a way to increase the robustness of entanglement, and as a result, increase the resistance of entanglement-based quantum communication to noise \cite{collins2002bell, Cerf2002, Thew2004, wiseman2007steering, Brunner2010}, see Refs.~\cite{Krenn:2017hz, erhard2018twisted} for an overview. \JL{These studies follow the work in Ref.~\cite{PhysRevLett.85.4418} that showed that violations of local realism are stronger for high-dimensional systems.}  In addition, encoding information in \mm{photonic} high-dimensional \mm{degrees-of-freedom such as transverse position/momentum and time/frequency} offers an increased \mm{information} bandwidth over qubit encoding, i.e.~greater than one bit of information per transmitted photon. This allows one to build efficient quantum networks and quantum cryptography systems that use the full information-carrying capacity of a photon \cite{Jha:2008wu, PhysRevA.85.060304, wang2018multidimensional, Bouchard2018experimental,Valencia:2019tc}.  

Several recent works on \mm{photonic} high-dimensional entanglement have demonstrated a qualitative advantage over qubit entanglement in terms of information-capacity and noise-robustness \cite{Ecker:2019ex, bavaresco2018measurements}. However, a careful analysis outlining the precise noise conditions and Hilbert space dimension where such an advantage can or cannot be found is still lacking. Here, we formulate these conditions in terms of easily measurable experimental parameters, and establish the exact noise bounds and Hilbert space dimension where high-dimensional entanglement provides an advantage for entanglement certification in the presence of noise, and where it does not.  We experimentally verify the predictions in the context of spatial entanglement generated in the few-photon limit, but the results are applicable to degrees of freedom where high-dimensional entanglement exists, e.g.~time-bin entanglement.  The results are also relevant to hyper and hybrid entanglement.


We follow a two-step approach to developing an operational noise model for \JL{photonic} high-dimensional entanglement: First, we show how the seemingly complex relationship between the sources of noise in the state, channel, and detection system can be distilled into one operational quantity---the signal-to-noise ratio $\mathcal{Q}$, which we refer to as the quantum contrast.  The quantity $\mathcal{Q}$ is simply the ratio of the coincidence counts to the accidental counts and can readily be established in any experiment. An analysis of the functional form of $\mathcal{Q}$ that we present will allow an experimenter to optimise their source, channel, and detector specifications in order to achieve the best noise performance (highest $\mathcal{Q}$) possible. Second, we test a series of contemporary entanglement measures and analyse their performance in the presence of noise. In all cases, we find that the performance of the system, as quantified by the ability to certify (high-dimensional) entanglement, can be accurately predicted by knowledge of the system dimension $d$ and only one experimentally measurable parameter: $\mathcal{Q}$.  

For high-dimensional photonic entanglement, our results show that separating the dimension in which one wishes to observe the entanglement and the dimension of the Hilbert space allows one to tolerate significantly more noise in an entanglement distribution system. However, depending on the entanglement certification method used, there is an optimum dimension where one finds the best possible noise performance. Interestingly, the largest increase in noise tolerance is obtained for only a modest increase in Hilbert space dimension, \mm{allowing the use of inefficient detectors or operation in extremely noisy environments.}


\section{Theoretical model}

\subsection{The relationship between $\mathcal{Q}$ and $p$ from the isotropic state}

An often-used model for describing noise in an entangled state combines a maximally entangled state $\ket{\phi}$ with a maximally mixed state $\hat{\mathds{I}}$ denoting white noise: 
\begin{align}\label{werner}
\hat{\rho} = p \ket{\phi}\bra{\phi}+\frac{(1-p)}{d^2}\hat{\mathds{I}}.  
\end{align}
This is often referred to as the isotropic state.  Here, $p$ refers to the probability that the state remains intact \cite{werner1989quantum}. One sees here that the resistance to noise will increase linearly with system dimension $d$---the threshold $p$ required to certify any entanglement is equal to $1/(d+1)$ \cite{collins2002bell}.  As the system dimension is increased, one can expect to overcome any amount of noise. For example, the threshold $p$ in $d = 2$ is $1/3$, whereas the threshold $p$ in $d = 10$ reduces to only $1/11$.   However, in any realistic scenario, $p$ turns out to be a dimension-dependent parameter that involves a complex interplay between noise attributed to the state, the channel, and the detection system.  

\begin{figure}[]
\centering
\includegraphics[]{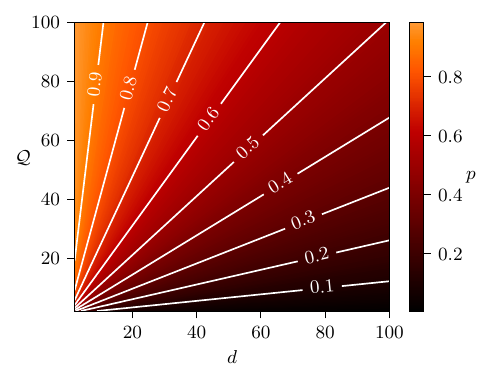}
\caption{The relationship between $\mathcal{Q}$, $d$ and $p$ from the isotropic state. The white lines indicate constant values of $p$.  We see that maintaining a constant value of $p$ as $d$ increases requires an increase to the signal-to-noise ratio $\mathcal{Q}$.}
\label{fig:werner}
\end{figure}

In order to realistically capture how noise translates in a high-dimensional system, here we use an operational definition of noise in terms of quantities easily measured in experiment. Our work focuses on the signal-to-noise ratio $\mathcal{Q}$ (defined in the next section), rather than $p$, but we can relate these two parameters to each other.  This is important as we can establish the link between theory and experiment is a simple fashion.  We can relate the $p$ from the isotropic state to $\mathcal{Q}$ and dimension $d$ through
\begin{align}
p(d,\mathcal{Q}) = \frac{\mathcal{Q}-1}{\mathcal{Q}-1+d}.\label{relation_p_qc_d}
\end{align}
Solving this for $\mathcal{Q}$, we see that for a fixed value of $p$ and as $d$ increases, it is necessary to increase $\mathcal{Q}$.    For example, to achieve $p = 1/3$ in $d = 2$ requires that that the signal to noise equal $\mathcal{Q} = 2$, whereas to achieve $p = 1/3$ in $d = 10$ requires that  $\mathcal{Q} = 6$.  This is illustrated in Fig.~\ref{fig:werner} where we see the relationship between $d$, $\mathcal{Q}$, and $p$.

\subsection{Definition of $\mathcal{Q}$}

We introduce a formalism that takes into account the common sources of noise in photonic systems: photon-pair generation probability in spontaneous parametric downconversion (SPDC) \JL{$\mu$} \cite{christ2011probing,yuen1976two, takeoka2015full,alibart2006photon,wasilewski2008statistics}, imperfect measurements with dark counts from the detector \JL{and} background counts from ambient light sources rolled into a term \JL{$n$} \cite{hadfield2009single,rohde2006modelling,takesue2007quantum}, and \JL{a term that accounts for any loss or non-unity collection efficiency $\eta$}.  \JL{The system  we consider uses single-photon detectors without the ability to perform photon-number detection.  Adding photon-number resolving detection would enable the detection of entanglement with more multi-photon pairs, but in this work we consider the most widely used case for two-photon entanglement.   }

\JL{We consider the measurement interval to be the time that a detector is active, and the quantities \mm{$\mu$ and $n$} are then normalised with respect to this time window.  In the case that $\mu \ll 1$, higher order multi-photon terms do not significantly contribute, and $\mu$ can be considered as the probability of generating a photon pair per detection event.   The value $n$ is the probability of detecting any noise event in the time window.  In general, $n$ and $\eta$ can be dimension-dependent, however, in our theoretical formalism we treat them as being \mm{independent as this represents the best-case scenario for noise-resistance.}}

\begin{figure}[]
\centering
\includegraphics[]{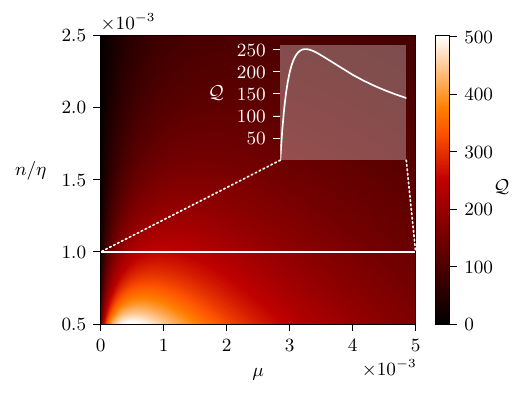}
\caption{The signal-to-noise ratio $\mathcal{Q}$ as a function of the noise-efficiency ratio $n/\eta$ and the photon pair generation rate $\mu$. Given a noise-efficiency ratio, there is an optimal generation rate that maximises the signal to noise.}
\label{fig:optimal}
\end{figure}

\JL{ The state that we consider is then
\begin{align}
\ket{\phi}=\prod_{j=1}^d \left[\sum_{m=0}^{+\infty} \left(\frac{\mu}{\mu+1}\right)^\frac{m}{2} \frac{\hat{a}_{1,j}^{\dag{m}}\hat{a}_{2,j}^{\dag{m}} }{m! \sqrt{1+\mu}}\right]\ket{{\rm{vac}}},\label{state}
\end{align}
where $\hat{a}_{1,j}^{\dag{m}}$ is the creation operator for $m$ photons in mode $j$.  Details of the derivation of Eq.~(\ref{state}) are given at the end of the paper.} We assume that the efficiencies and noise levels in each channel are the same, so the \JL{probability} of two-photon coincidences and accidentals \JL{per time window is given by $\rm{p_{coi}^{(j,k)}}=\eta^2\delta_{j,k}{\mu}(1+{\mu}) + (n+\eta{\mu})^2$, where $j$ and $k$ are indices for the different modes of the photons, and $1 \leq j, k \leq d$, with $d$ equal to the dimension of the Hilbert space.  
We define the signal-to-noise ratio or the quantum contrast $\mathcal{Q}$ as the ratio of coincidences} \JL{${\rm p_{coi}^{(j,j)}}$ }to accidentals \JL{${\rm p_{coi}^{(j, k\neq j)}}$}
\begin{align}
\mathcal{Q}=1+\frac{\mu(1+\mu)}{(\frac{n}{\eta}+{\mu})^2}.\label{qc_0_main}
\end{align}
\mm{The quantum contrast can be optimised by an appropriate choice of the pair generation rate $\mu$ based on the noise parameters $\eta$ and $n$, and achieves a maximum value of $\mathcal{Q}_{\rm max} = 1+\eta^2/[4 n (\eta-n)]$, when the pair generation rate is set to $\mu = n/(\eta-2n)$, see Fig.~\ref{fig:optimal}.}

\subsection{Assumptions of the model}

There are two important assumptions in our analytic theory: first, the dimension of the state Eq.~(\ref{state}) can be increased without any penalty; and second, the amplitudes of the mode coefficients are equal, i.e.~the state is maximally entangled.   These assumptions greatly simplify the theoretical description and provide an upper bound to noise resistance. In any physical system, however, there will be an upper limit to the size of the space, and real sources often require entanglement concentration in order to achieve a flat state \cite{dada2011experimental}.

\begin{figure}[]
\begin{subfigure}
  \centering
  \includegraphics[trim={0 0.3cm 0 0.2cm}, clip]{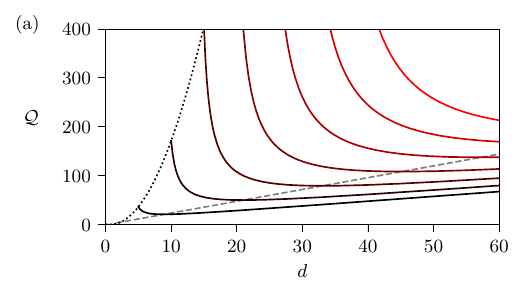}  
  \label{fig:dataA}
\end{subfigure}
\begin{subfigure}
  \centering
  \includegraphics[trim={0 0.3cm 0 0.2cm}, clip]{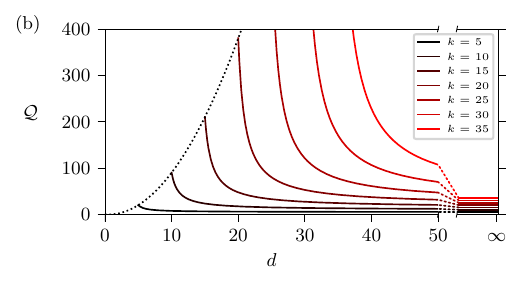}  
  \label{fig:dataB}
\end{subfigure}
\caption{(a). The required quantum contrast $\mathcal{Q}$ vs.~dimension $d$ for the two-MUB witness.  The dotted line represents the necessary $\mathcal{Q}$ for $k = d$ dimensional entanglement. The dashed line represents the optimal dimension for measuring  $k-$dimensional entanglement. (b). Required quantum contrast $\mathcal{Q}$ vs.~dimension $d$ for various values of $k$ for the $d+1$-MUB witness.   The contrast necessary for $d$-dimensional entanglement in a $d$-dimensional space is indicated by the dotted line and is equal to $d^2-d$;  the contrast necessary for $k$-dimensional entanglement in an infinite space is equal to $k$, indicated in the bottom right-hand side of the figure.  }
\label{fig:theory}
\end{figure}

\subsection{Entanglement verification via two mutually unbiased bases}

Recently, it was shown that measurements in two mutually unbiased bases (MUBs) are sufficient to lower bound the fidelity of a state and certify high-dimensional entanglement \cite{bavaresco2018measurements}. \mm{If the fidelity $F(\rho,\Phi)$ of the state $\rho$ with respect to a $d$-dimensional target state $\Phi$ is greater than $(k-1)/d$, $\rho$ is entangled in at least $k$ dimensions, i.e.~it has Schmidt rank of at least $k$.} As we consider a state with a uniform modal distribution, the quantum contrast is \mm{basis-independent. The two-MUB witness can then be used to lower bound the achievable fidelity $\tilde{F}$ of a $d$-dimensional state with quantum contrast $Q$:}
\begin{align}
&\tilde{F}(\rho,\Phi) \geq  \frac{\mathcal{Q} -d+1}{\mathcal{Q}+d-1}.   \label{mub2bases_qc_1_main}
\end{align}
%
\mm{Consequently, in order to certify $k$-dimensional entanglement, the quantum contrast must satisfy}
\begin{align}
\mathcal{Q}>&\frac{(d-1)(d+k-1)}{d-k+1}.\label{mub2bases_qc_2_main}
\end{align}
We see that as $k$ is increased, the minimum contrast that is required increases.  Importantly, as shown in Fig.~\ref{fig:theory}a, we see that in order to certify $k$-dimensional entanglement, a minimum quantum contrast of $\mathcal{Q_{\rm opt}} = 3 k+2 \sqrt{2} \sqrt{(k-2) (k-1)}-4$ is necessary, which is obtained when the dimension of the Hilbert space we are working in is set to $d_{\rm opt}= [\sqrt{2} \sqrt{k^2-3 k+2}+k-1]$.

We see that $d_{\rm opt}$ is reached by increasing $k$ by approximately $(1+\sqrt{2}) \approx 2.41$ times, at which we have significantly increased our system's ability to tolerate noise.  The gain can be quantified as the ratio of $\mathcal{Q}$ when $d = k$ and when $d = d_{\rm opt}$.  This ratio is approximately $(2/(3+2\sqrt{2}))k \approx 0.343 k$, so if $k = 1000$, increasing $d$ by 2.41 times reduces $\mathcal{Q}$ by 343. From an experimental perspective, this decrease in $\mathcal{Q}$ permits the use of a detector with a significantly lower detection efficiency/channel loss, or a higher dark/background count rate.

\subsection{Example for an EMCCD camera}

As an example of the gains this provides, consider a multi-outcome detector, such as a single-photon detector array or EMCCD camera \cite{edgar2012imaging, defienne2018adaptive}.  These detectors are commonly used for measurements of spatial entanglement as they allow multi-outcome measurements in two MUBs, i.e.~ the position and momentum bases.  Certifying $k = 1000$ entanglement in a $d = 1000$ space using two MUBs requires a signal-to-noise of approximately $2\times10^6$. Achieving this signal-to-noise requires, for example, a noise per detection event of $n = 1\times10^{-7}$ and an efficiency of $\eta = 80\%$.  Increasing the dimension of the space by a factor 2.41, (e.g.~moving from a 32 by 32 array of pixels to a 49 by 49 grid), reduces the necessary $\mathcal{Q}$ by $\approx$343 times, and, for example, the allowable detector noise can increase two orders of magnitude to $n = 1\times10^{-5}$ while the efficiency simultaneously drops to $23\%$.

\subsection{Entanglement verification via all mutually unbiased bases}

\mm{Measurements in all $d+1$ MUBs allow one to calculate the exact fidelity while also providing better noise performance \cite{bavaresco2018measurements}. In this case, the fidelity is given by}
\begin{align}\label{F_all_MUBs}
\tilde{F}(\rho,\Phi) = \frac{\mathcal{Q}+\frac{1}{d}-1}{\mathcal{Q}+d-1}.
\end{align}
%
\mm{Then, for a $d$-dimensional system with quantum contrast $Q$, one can certify an entanglement dimensionality of}
\begin{align}
k < \frac{(d+1) \mathcal{Q}}{d+\mathcal{Q}-1}.\label{we2_qc}
\end{align}
\JL{We see here that knowledge of $\mathcal{Q}$ provides an upper limit to the number of entangled dimensions. In any experiment an estimate of $\mathcal{Q}$ can be established easily, and therefore, $k$ can be predicted in a fast and accurate manner. }

\mm{As in the two-MUB case, entanglement certification in increasing $k=d$ dimensions necessitates an accompanying increase in the required quantum contrast (Fig.~\ref{fig:theory}b).} \JL{However, significant reductions in the required quantum contrast are achieved when we allow the dimension of the Hilbert space to increase with respect to the dimension of the entanglement, i.e.~if $k < d$. In the limit that $d \rightarrow \infty$, we see that the number of entangled dimensions $k \rightarrow \mathcal{Q}$.} It follows that if we measure a system with a contrast of $\mathcal{Q}$, it can have at most $\lfloor \mathcal{Q} \rfloor$ dimensions of entanglement. Consequently, the minimum contrast required for verification of $k$ dimensions of entanglement is equal to $k$.


\section{Numerical and Experimental Results}

\begin{table*}[!htbp]
\centering
\begin{tabular}{|c||c|c|c||c|c||c|c|}
\hline
{} &  \multicolumn{3}{c||}{Measured data from Ref.~\cite{bavaresco2018measurements} } & \multicolumn{2}{c||}{Predictions based on $\mathcal{Q}_{\rm exp}$}
& \multicolumn{2}{|c|}{Optimal conditions}\\
\hline
\hline
$d$   & $\tilde{F}_{{\rm exp}}$   & $k_{\rm{exp}}$   & $\mathcal{Q}_{\rm exp}$ & $\tilde{F}_{\rm pred}$   & $k_{\rm pred}$ & $d_{\rm opt}$ & $ \mathcal{Q}_{\rm opt}$\\
3   & 91.5$\pm 0.4\%$ & 3 & 71$\pm$1 & 94.5$\pm0.1$\% & 3 & 4  & 9.0 \\
5   & 89.9$\pm 0.4\%$ & 5 & 70$\pm$5 & 89.2$\pm0.7$\% & 5 & 9  & 20.8 \\
7   & 84.2$\pm 0.5\%$ & 6 & 68$\pm$4 & 83.9$\pm0.9$\% & 6 & 14 & 32.5\\
11  & 74.8$\pm 0.4\%$ & 9 & 81$\pm$2 & 78.1$\pm0.5$\% & 9 & 23 & 55.8\\
\hline
\end{tabular}
\caption{Comparison of the results from Ref.~\cite{bavaresco2018measurements} against the predictions of this work. The data from \cite{bavaresco2018measurements} was evaluated to calculate an average signal-to-noise ratio for the two MUBs in each of the dimensions.  These $\mathcal{Q}_{\rm exp}$ values are then used to predict the fidelity $\tilde{F}_{\rm pred}$ and dimension $k_{\rm pred}$ of the entanglement.  For every dimension $d$, the experimentally observed $k_{\rm exp}$ and the predicted $k_{\rm pred}$ are equal to each other, thus confirming that the system performance can accurately be predicted with knowledge of $\mathcal{Q}$.}\label{table1}
\end{table*}

\subsection{Analysis of data from Ref.~\cite{bavaresco2018measurements}}

In order to verify that knowledge of the signal-to-noise ratio provides an accurate estimate of system performance when using two MUBS, we re-analyzed the data presented in Ref.~\cite{bavaresco2018measurements} to calculate $\mathcal{Q}_{\rm exp}$ for  the measurements in each of the MUBs and used this to predict the system performance, see Table \ref{table1}.  In each dimension ($d$ = 3, 5, 7, and 11), we see that knowledge of $\mathcal{Q}_{\rm exp}$ gives an estimate of ($k_{\rm pred}$ = 3, 5, 6, and 9) respectively, which are the exact values of dimension obtained.  This confirms that knowledge of a single, easily obtainable, experimentally measured parameter $\mathcal{Q_{\rm exp}}$ provides a very accurate prediction of system performance.

\subsection{Experimental verification}

\begin{figure}[]
\centering
\includegraphics[]{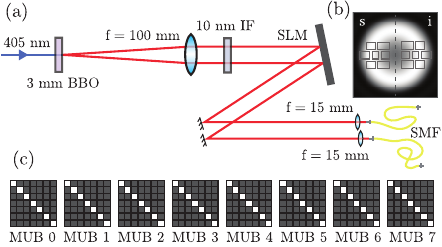}
\caption{\JL{(a)~Schematic of experiment to generate and measure high-dimensional entanglement in the pixel basis. BBO = Beta barium borate; IF = interference filter; SLM = spatial light modulator; SMF = single-mode fibre. (b)~Illustration of the modes on the SLM used to detect high-dimensional entanglement.  The left half of the SLM is used to detect the signal (s); the right half is used to detect the idler (i). Each small box represents a mode in the computational MUB.  (c) In seven dimensions, there are eight MUBS, and MUB 0 is the computational basis. The signal-to-noise ratio $\mathcal{Q}$ is assumed to be constant across all MUBs in the theory and we calculate the average $\mathcal{Q}$ across the MUBs in the experiment.  }}
\label{fig:setup}
\end{figure}

\JL{To investigate the predictions of our theory, we perform measurements on the high-dimensional entangled state produced by SPDC, see Fig.~\ref{fig:setup}(a).  The measurements are performed using the spatial degree of freedom in all MUBs in 3, 5, and 7 dimensions with \mm{average quantum contrasts ranging from $\mathcal{Q}=5$ to 40}. $\mathcal{Q}$ is varied by adjusting the laser pump power and ambient lighting conditions in the lab and is calculated as the average of the signal-to-noise ratios of each of the MUBs. }

All the measurements are conducted in the ``pixel" bases \cite{valencia2020}, where the computational basis uses the standard anti-correlations in photon momenta that are are observed in the far-field of SPDC.  Fig.~\ref{fig:setup}(b) indicates the intensity distribution of the far-field of the SPDC together with the areas integrated for the measurement modes.   We use one SLM, where the left half used for the signal photon, and the right half is used for the idler photon.  For the computational basis measurements (MUB 0), a single area of the signal and idler is active at any one time; for all the other MUB measurements, all elements are active all times with the phase of each area appropriately controlled.  The coincidence matrices for all the MUBS, and thus the joint probabilities, are measured by scanning through the appropriate signal and idler modes on the SLM. 
 
\begin{figure}[]
\begin{subfigure}
  \centering
  \includegraphics[]{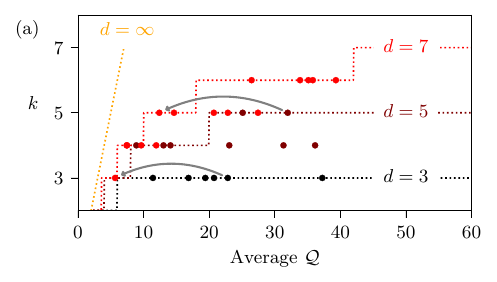}  
  \label{fig:dataA}
\end{subfigure}
\begin{subfigure}
  \centering
  \includegraphics[]{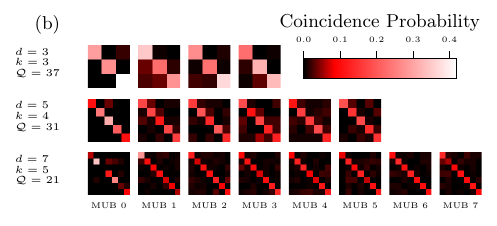}  
  \label{fig:dataB}
\end{subfigure}
\caption{\JL{(a) Experimental certification of $k$-dimensional entanglement for signal-to-noise ratios ranging from $\mathcal{Q}=5$ to 40 for $d = 3, 5,$ and $7$.  The dotted lines are the analytical theory, the solid points are the measured values of $k$.  The two grey arrows represent the lowering of $\mathcal{Q}$ from 32 to 12 for 5-dimensional entanglement and 23 to 6 for 3-dimensional entanglement.  In both cases, provided that $d$ is increased, high-dimensional entanglement remains even if the signal to noise falls. (b) The measurement matrices showing the joint measurement  probabilities for three sample data sets.  }}  
\label{fig:exp}
\end{figure}

\JL{The data are analysed to calculate the lower bound to the fidelity and thus a value of $k$.  
Figure \ref{fig:exp}a shows experimental evidence of two of the results of this work: first, that knowledge of $\mathcal{Q}$ and $d$ provides an accurate prediction of system performance with respect to entanglement measures; and second, that \mm{increasing the operational dimension allows one to tolerate larger amounts of noise for the same (or larger) certified entanglement dimensionality.}} \JL{The measured values of $k$ plotted in Fig.~\ref{fig:exp} are observed to be very close to, but not above, the upper bound predictions of Eq.~\ref{we2_qc}.  We observe that entanglement in larger dimensions ($k$ increases) can be achieved by increasing $d$ while tolerating a lower signal-to-noise ($\mathcal{Q}$ drops). }

\begin{figure}[]
\centering
\includegraphics[]{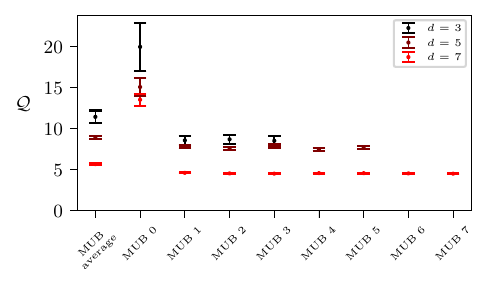}
\caption{The measured signal-to-noise ratios for different MUBs in different dimensions.  This is a selection of the data used that is typical of the observed trends. This shows that MUB 0 has a higher $\mathcal{Q}$ value than the other MUBs, which are all measured to be close to each other.  The MUB average is the value that is used for the predictions of system performance. }
\label{fig:Q_error}
\end{figure}

Figure \ref{fig:Q_error} shows the variation of the measured signal to noise ratios for different MUBs in different dimensions.  We see that the $\mathcal{Q}$ value for the computational MUB (MUB 0) is consistently higher than the values for the other MUBs, which are observed to be very close to each other.   We believe that this is a systematic error introduced by the SLM-based measurement method, rather than anything inherently in the generated state.  Despite this systematic error, the predicted performance of system with regards to all entanglement metrics still performs exceptionally well when using the average value of $\mathcal{Q}$ across all the different MUBs.

The the signal-to-noise values for each MUB are calculated from the appropriate coincidence probability matrix, Fig.~\ref{fig:exp}(b).   For one MUB, it is calculated as the average of the diagonal elements of the matrix divided by the average of the off-diagonal elements of the matrix.   We chose this approach to measure the most accurate value of $\mathcal{Q}$, however, a very fast estimate of $\mathcal{Q}$ can be obtained by measuring the coincidence rates for only one diagonal and one off-diagonal element of any of the coincidence matrices.    The predicted system performance when using this fast method will depend on any error between the estimated $\mathcal{Q}$ and the true value.

\subsection{Numerical verification}

Our theory considers maximally entangled states as this assumption considerably simplifies the theory.   To take into account the effects of a reduced bandwidth, we perform numerical simulations on a finite bandwidth state.  The width of the state is controlled by setting the probability amplitude of the single photons to have a Gaussian distribution.  We use $\sigma = 2, 4, 10$ and 100000.   A necessary consequence of a limited bandwidth is there exists crosstalk in MUBs other than the computational basis.  These simulations therefore enable us to investigate realistic quantum sources that exhibit common types of measurement error.  We introduce noise into the simulated data and then calculate an average $\mathcal{Q}$ value over all the MUBs.

\begin{figure*}
\centering
\includegraphics[trim=100 160 40 85, clip,  width = 12 cm]{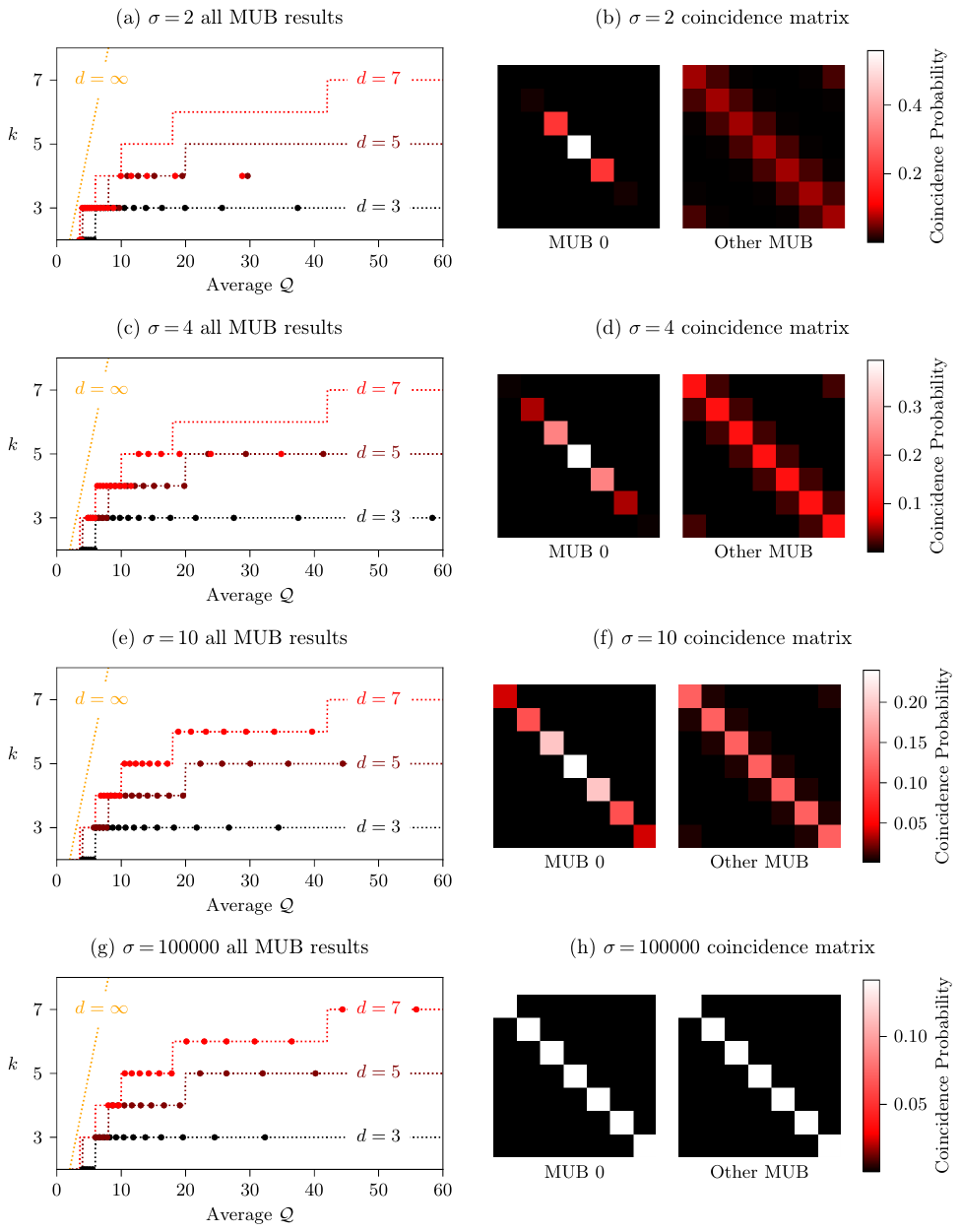}
\caption{Numerical simulation of average quantum contrast vs.~entangled dimension for a range of finite-width states.  Each point represents one numerical simulation, and the dotted lines are the analytical thoery.  Increasing the size of the space $d$ continues to provide noise resistance, but only up to certain point governed by the width of the state.  These numerical simulations were performed by setting the width of the probability amplitude of the single photon to have a Gaussian envelope of width $\sigma$ for a wide range of signal to noise values.  (a), (c), (e) and (g) all show the average $\mathcal{Q}$ vs.~$k$ for $\sigma$ = 2, 4, 10 and 100000, respectively. (b), (d), (f) and (h) are samples of the coincidence matrices for $\sigma$ = 2, 4, 10, and 100000 respectively, showing the finite width of the state and the influence this has on cross-talk measurements in the MUBs. The $\sigma$ = 100000 case is a close approximation to the state that we consider in the analytical theory, and we recover the same results for this case.}
\label{fig:NumData}
\end{figure*}

Figure \ref{fig:NumData} shows the results for numerical simulations of $k$-dimensional entanglement certification for a state with a finite number of entangled modes.   Despite the finite widths, the two main results of our work are still observed: the average $\mathcal{Q}$ is a very good indicator of system performance, and increasing the size of the space ($d$) enables the certification of entanglement in larger dimensions with lower signal-to-noise.   The difference from the analytical theory, which provides an upper bound, is that the size of the space cannot be increased indefinitely.  One finds that there is an upper limit to the advantage of increasing $d$.   The $\sigma = 100000$ result, which represents a state with a very large bandwidth, is a very close match to the analytical theory.

\section{Additional Entanglement Measures}

\subsection{Conditional Entropy} 

\begin{figure}[]
\centering
\includegraphics[]{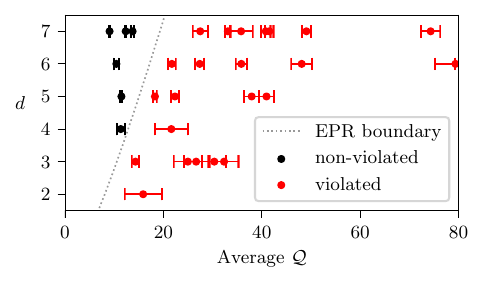}
\caption{\JL{Certification of EPR steering in high dimensions for a range of signal-to-noise ratios.  Measurements in two MUBs for a range of signal-to-noise ratios were taken and the data tested against the EPR steering criteria.  The data are plotted as either red or black points depending on whether the EPR criteria is violated or not.  The dotted line is the boundary that is calculated by solving Eq.~\ref{entrop_cri0}.  The derived bound, based only on $\mathcal{Q}$ and $d$, accurately separates the experimentally measured data.  }}
\label{fig:EPR}
\end{figure}

We also analyse \mm{the noise dependance of} conditional entropies commonly used in the confirmation of \mm{EPR steering} \cite{Reid2009,leach2010quantum, Schneeloch2013}. EPR \JL{steering} can be confirmed if the measurements in two \mm{MUBs} violate the inequality $
H_1(X|Y) + H_2(X|Y) \ge \log_2 d$, where $H_1(X|Y)$ and $H_2(X|Y)$ are the conditional entropies in each of the bases \cite{leach2012secure}. After we consider the sum of the conditional entropies \JL{in two bases}, we find that 
\begin{align}
\log_2(\mathcal{Q}+d-1)-\frac{\mathcal{Q}}{\mathcal{Q}+d-1}\log_2\mathcal{Q}-\frac{1}{2}\log_2d < 0.\label{entrop_cri0}
\end{align}
There is no analytical solution to this, but numerical solutions show that \mm{the necessary quantum contrast $\mathcal{Q}$ required to demonstrate EPR steering increases linearly as a function of Hilbert space dimension $d$, see dotted line in Fig.~\ref{fig:EPR}.} \mm{Experimental data for a range of dimensions showing whether the steering criterion is violated (red) or not (black) is also plotted as a function of quantum contrast. Clear agreement is seen with the predictions of Eq.~\ref{entrop_cri0}, which demonstrates that knowledge of the signal-to-noise ratio $\mathcal{Q}$ accurately predicts whether the EPR steering criterion will be violated.}


\subsection{High-dimensional non-locality based on the CGLMP inequality}

Finally, we consider the CGLMP inequality \cite{collins2002bell,dada2011experimental} that can be used for establishing non-locality.  Local hidden variable theories are consistent with $S_d \leq 2$, whereas quantum mechanics permits a violation of this inequality.  In their work, CGLMP showed that maximally entangled quantum systems in high dimensions can achieve a theoretical value of $S_d(QM)$, which would lead to a violation of the inequality \cite{collins2002bell}.  However, in any experimental verification, the achievable $S_d$ is modified by imperfections in the system.  \mm{Under the fair-sampling assumption,} the maximum value of $S_d$ \mm{given by}
\begin{align}
\tilde{S}_d=& \frac{\mathcal{Q}-1}{\mathcal{Q}-1+d} S_d(QM).\label{cglmp4_main}
\end{align}
\JL{Therefore, in order to violate the local hidden variable inequality, the number of dimensions must satisfy
\begin{align}
d < \frac{(\mathcal{Q}-1)(S_d(QM)-2)}{2}.\label{cglmp4}
\end{align}
For large values of $d$, $S_d(QM) \approx 3$, and the upper bound for $d$ can be approximated by $d < (\mathcal{Q}-1)/2$.}   Solving for $\mathcal{Q}$, we see that to violate the CGLMP inequality in $d$ dimensions, we must have a signal-to-noise ratio greater than $2d+1$, e.g.~violating the inequality in 10 dimensions requires that $\mathcal{Q}$ exceed $\approx$21.

As in the case of the fidelity witness, the separation of $k$ and $d$ may provide \mm{a noise advantage in the form of a decreased signal-to-noise threshold.} It will also be interesting to consider the implications of this work for other Bell\mm{-like} inequalities, such as the one developed for maximally entangled states \cite{salavrakos2017bell}.

\section{Discussion and conclusion}

This work serves to answer a simple question:~``{\it Are \mm{photonic} high-dimensional \mm{entangled} states robust to noise?}"  In reality, the answer to this question is more nuanced than a simple ``{\it yes}" or ``{\it no}", and our results provide a clear demonstration of the advantages and disadvantages.  When we consider the \mm{maximally entangled} state, we find that the total noise that the state can tolerate increases as the dimension of the state increases, and therefore, it is obvious to claim that such states are robust to noise.  However, a counter argument can be put forward when we recognize that the  experimental signal-to-noise tends to increase as we increase the dimension of the state in which we wish to observe entanglement or non-locality.

 Significant benefits for entanglement certification emerge when we see that the dimension of the state in which we wish to establish entanglement $k$ and the dimension of the \mm{operational Hilbert space} $d$ are not required to be the same. The signal-to-noise requirements for $k$-dimensional entanglement decrease as both $d$ and the total number of mutually unbiased bases used in the measurement increases.  Thus, by modest increases to the dimensionality of a system or through the use of judiciously chosen additional measurements, one can significantly increase the tolerance of high-dimensional entanglement to common sources of noise, such as loss and background counts.
 
 By distilling the information about the performance of a \mm{photonic} high-dimensional entanglement system into a single operational parameter, we provide a powerful tool that links theory and experiment, allowing us to accurately predict system performance and choose an optimum entanglement measurement strategy.  This work demonstrates that high-dimensional entanglement has the potential to push the boundaries of entanglement-based quantum communication systems, bringing them from the confines of laboratory proof-of-principle demonstrations to the realm of practical, real-world implementations under extreme conditions of noise.


\section{Acknowledgments}
We thank M.~Huber and N.~Friis for helpful discussions.
M.M.~and N.H.~acknowledge support from the QuantERA ERA-NET Co-fund (FWF Project I3773-N36) and the UK Engineering and Physical Sciences Research Council (EPSRC) (EP/P024114/1). J.L.~and F.Z.~acknowledge support from the UK EPSRC (EP/M01326X/1).


%

\subsection{Generation of high-dimensional entanglement through nonlinear optical processing}

The Hamiltonian of the electromagnetic field in the interaction picture in nonlinear optical media is given by

\begin{align}  
\hat{\rm{H}}'=&e^{i\frac{\hat{H}_0}{\hbar}t}\hat{\rm{H}}_{\rm{NL}}e^{-i\frac{\hat{H}_0}{\hbar}t}\label{H'} \nonumber \\ 
=&\sum_{j_1,j_2,j_3}\hat{a}_{j_1}^{\dag}\hat{a}_{j_2}^{\dag}\hat{a}_{j_3}{\rm{sinc}}[\frac{1}{2}\bm{L}\cdot(\bm{k}_{j_1}+\bm{k}_{j_2}-\bm{k}_{j_3})]\nonumber \\
&{e}^{i(\omega_1+\omega_2-\omega_3)t}\Gamma_{j_1,j_2,j_3}+{\rm{H.c.}}
\end{align}
$\hat{a}_{j_1}^{\dag}\hat{a}_{j_2}^{\dag}$ indicates photon-pair generation events;  $H.c$ is the Hermitian conjugate; $\hat{a}_{j_3}$ represents the annihilation of a pump photon, showing that photon pair generation is proportional to the pump power; and $|\bm{L}|$ is the thickness of the crystal.   $\Gamma_{j_1,j_2,j_3}$ is a function associated with $\bm{\chi}^{(2)}$, $|\bm{L}|$, $\hbar$, $\epsilon_j$, and the effective cross-section area of the optical modes. 

Furthermore, if $\omega_1+\omega_2-\omega_3{\neq}0$, the integral of ${e}^{i(\omega_1+\omega_2-\omega_3)t}$ in the time domain will  vanish, ensuring energy conservation. In addition, the ${\rm{sinc}}\{\frac{1}{2}\bm{L}\cdot[\bm{k}_{j_1}+\bm{k}_{j_2}-\bm{k}_{j_3}]\}$ term is from integration in the spatial degrees of freedom, ensuring momentum conservation. Based on these two conditions, there is only one independent variable $j$, so that we have
\begin{align}
\hat{\rm{H}}'=&\sum_{j=1}^d\gamma_j\hat{a}_{1,j}^{\dag}\hat{a}_{2,j}^{\dag}\hat{a}_{3}+{\rm{H.c.}}\label{hh2}
\end{align}
where $d$ is the dimension of entanglement modes, $\gamma_j$ is the reduction of $\Gamma_{j_1,j_2,j_3}$ in Eq.~($12$), $\hat{a}_{1,j}^{\dag}$,  $\hat{a}_{2,j}^{\dag}$ is the creation operators for signal and idler photons, respectively, and $\hat{a}_{3}$ represents the annihilation operator of a pump photon. Usually the pump light includes a very large number of photons. Hence, the order of $\hat{a}_{3}$ and $\hat{a}_{3}^{\dag}$ is no longer important. Then $\hat{a}_{3}$ and $\hat{a}_{3}^{\dag}$ can be replaced by $\alpha_p$ and $\alpha_p^{*}$. The final state is
\begin{align}
\ket{\phi}=&\exp\{\sum_{j=1}^d\frac{i}{\hbar}t\gamma_j\alpha_p\hat{a}_{1,j}^{\dag}\hat{a}_{2,j}^{\dag}-{\rm{H.c.}}\}\ket{{\rm{vac}}}\nonumber\\
=&\prod_{j=1}^d\exp\{\frac{i}{\hbar}t\gamma_j\alpha_p\hat{a}_{1,j}^{\dag}\hat{a}_{2,j}^{\dag}-{\rm{H.c.}}\}\ket{{\rm{vac}}}\nonumber\\
=&\prod_{j=1}^d[\sqrt{1-|f_j|^2}\sum_{m=0}^{+\infty}\frac{1}{m!}f_j^k\hat{a}_{1,j}^{\dag{m}}\hat{a}_{2,j}^{\dag{m}}]\ket{{\rm{vac}}}\label{hh3}\nonumber\\
=&\prod_{j=1}^d \left[\sum_{m=0}^{+\infty} \left(\frac{\mu}{\mu+1}\right)^\frac{m}{2} \frac{\hat{a}_{1,j}^{\dag{m}}\hat{a}_{2,j}^{\dag{m}} }{m! \sqrt{1+\mu}}\right]\ket{{\rm{vac}}}
\end{align}
where $t$ is time, $f_j$ is associated with $\hbar$, $t$, $\gamma_j$ and $\alpha_p$, showing the intensity of correlated photon pairs in each mode. $\ket{{\rm{vac}}}$ indicates the initial state which is the vacuum state for signal and idler photons. 

The maximally entanglement state is widely used, which requires $f_j={\rm{constant}}$. If all $f_j$ are not equal, entanglement concentration can be used. We can replace all $f_j$ with a constant so that   
\begin{align}
|f_j|^2=&\frac{{\mu}}{{\mu}+1}\label{ff}
\end{align}
where $d$ is the dimension of entanglement, and ${\mu}$ probability of generating a photon pair per detection event.

\end{document}